\documentclass[twocolumn,showpacs,prb,aps,amsmath,amssymb,superscriptaddress]{revtex4}
\usepackage{amsmath}
\usepackage{graphicx}
\usepackage{rotating}

\def\bbbc{{\mathchoice {\setbox0=\hbox{$\displaystyle\rm C$}\hbox{\hbox
to0pt{\kern0.4\wd0\vrule height0.9\ht0\hss}\box0}}
{\setbox0=\hbox{$\textstyle\rm C$}\hbox{\hbox
to0pt{\kern0.4\wd0\vrule height0.9\ht0\hss}\box0}}
{\setbox0=\hbox{$\scriptstyle\rm C$}\hbox{\hbox
to0pt{\kern0.4\wd0\vrule height0.9\ht0\hss}\box0}}
{\setbox0=\hbox{$\scriptscriptstyle\rm C$}\hbox{\hbox
to0pt{\kern0.4\wd0\vrule height0.9\ht0\hss}\box0}}}}

 \usepackage{color}
\usepackage{here}
\definecolor{DarkBlue}{rgb}{0.1,0.1,0.5}
\definecolor{Red}{rgb}{0.9,0.0,0.1}
\definecolor{Green}{rgb}{0.0,0.99,0.0}

\newcommand{\beq}{\begin{eqnarray}}
\newcommand{\eeq}{\end{eqnarray}}

\newcommand{\bk}{{\bf k}}

\newcommand{\beqa}{\begin{eqnarray}}
\newcommand{\eeqa}{\end{eqnarray}}

\begin{document}

\title{Bogoliubov Angle, Particle-Hole Mixture
 and Angular Resolved Photoemission Spectroscopy in Superconductors.}

\author{Alexander V. Balatsky}
\affiliation{Theoretical Division and  Center for Nonlinear Studies,
Center for Integrated Nanotechnologies, Los Alamos National
Laboratory, Los Alamos, New Mexico 87545, USA}

\author{W. S. Lee}
\affiliation {Department of Physics, Applied Physics, and Stanford Synchrotron Radiation Laboratory, Stanford University, Stanford, CA 94305}

\author{Z. X. Shen}
\affiliation {Department of Physics, Applied Physics, and Stanford Synchrotron Radiation Laboratory, Stanford University, Stanford, CA 94305}

\date{Printed \today }

\begin{abstract}
Superconducting excitations ---Bogoliubov quasiparticles ---
 are the quantum mechanical mixture of negatively
charged electron (-e)  and positively charged hole (+e).   We
propose a new observable  for Angular Resolved Photoemission
Spectroscopy  (ARPES) studies that is the manifestation of the
particle-hole entanglement of the superconducting quasiparticles. We
call this observable a {\em Bogoliubov angle}. This angle measures
the relative weight of particle and hole amplitude in the
superconducting (Bogoliubov) quasiparticle. We show how this
quantity can be measured  by comparing the ratio of spectral
intensities at positive and negative energies.
\end{abstract}
\pacs{Pacs Numbers: }

\maketitle

\vspace*{-0.4cm}

\columnseprule 0pt

\narrowtext \vspace*{-0.5cm}

i) {\em BA and particle-hole mixture in paired states}. The analog
of the conduction electrons in the superconductors are the
quasiparticles. Unlike electrons, the superconducting quasiparticles
do not carry definite charge. The same quantum mechanical dualism
that allows electron to be at the same in two states is at play when
one considers the Bogoliubov quasiparticles in superconducting
state: the quasiparticle is a coherent combination of an electron
and its absence (``hole''). Particle-hole dualism of quasiparticles
is responsible for a variety of profound phenomena in
superconducting state such as Andreev reflection, the particle-hole
conversion process that is only possible in superconductor.

\begin{figure}[htb]
\begin{center}
\includegraphics[width= 2 in]{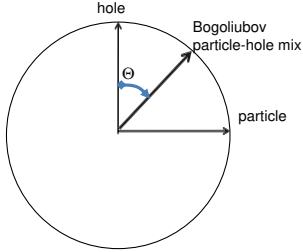}
\caption{ \label{scangle1} (Color online) Circle parametrizing
Bogoliubov admixture angle is shown. For $\Theta = 0, \pi/2$ the
mixture reduces to purely hole-like and particle-like state. At
arbitrary angle one deals with true Bogoliubov quasiparticles.}
\end{center}
\end{figure}

In this paper we introduce a quantity that parametrizes the mixture
in terms of an angle, we call this angle a {\em Bogoliubov angle}
(BA), see Fig.~\ref{scangle1}. We discuss how  ARPES measurements
allow one to visualize the Bogoliubov angle and thus to reveal
particle hole dualism. Here we introduce BA for ARPES in a similar
way as it has been introduced in STM
 \cite{Fujita:2007}.

\begin{figure}[htb]
\begin{center}
\includegraphics[width=2 in]{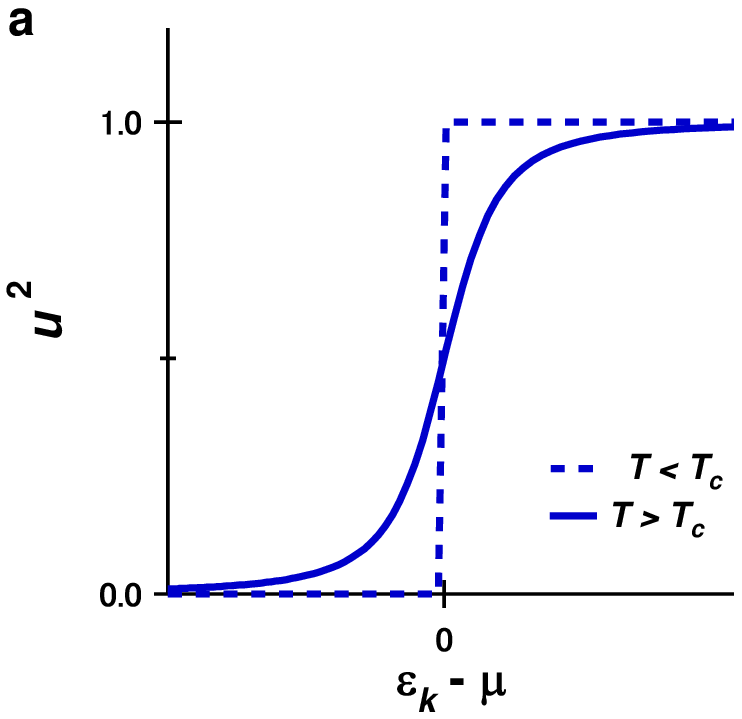}
\includegraphics[width=2 in]{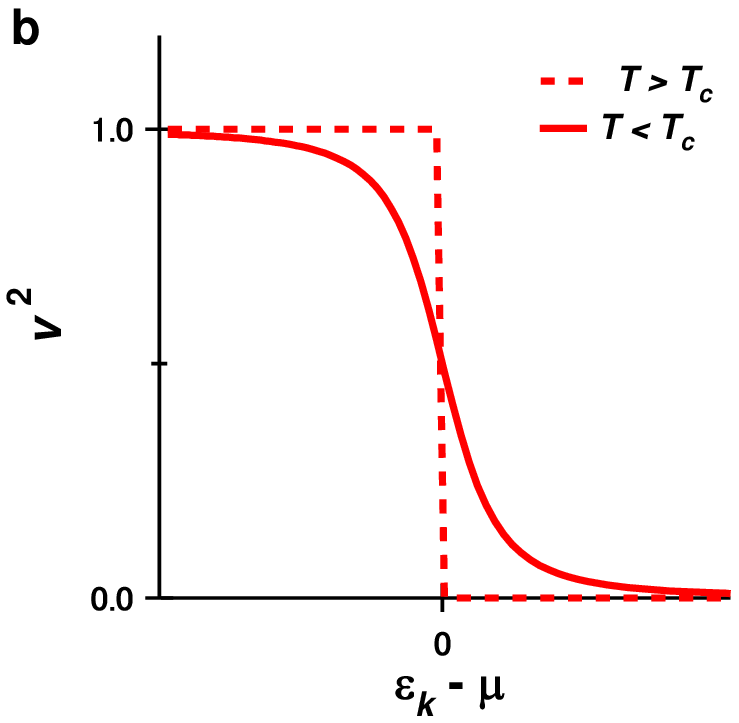}
\includegraphics[width=2 in]{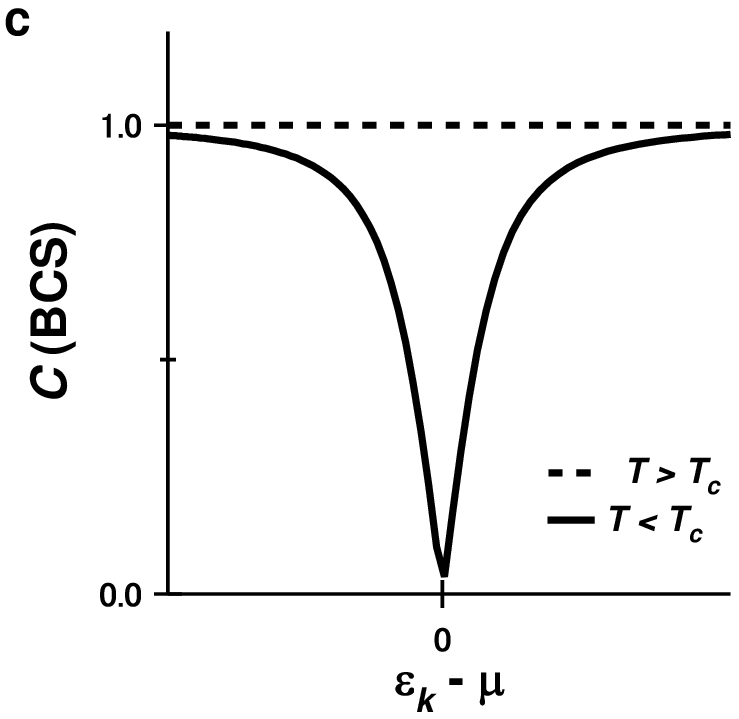}
\caption{\label{concurrence1} (Color online) BCS coherence factors
$u^2(\bk), v^2(\bk)$ are shown as functions of energy. The function
$C(\bk) = |u^2(\bk)- v^2(\bk)| = |\cos 2\Theta(\bk)|$,  shows
substantial departures from unity only in the energy range on the
scale of the gap $\Delta$ near the Fermi energy, where there are
substantial pairing correlations.}
\end{center}
\end{figure}

To illustrate the point about BA, we can look at the textbook BCS
case first.  We will show that convenient definition of BA is :
\beqa \Theta_k = \arctan [(\frac{|u(\bk)|^2}{|v(\bk)|^2})^{1/2}],
\label{EQ:BCSBA} \eeqa with the conventional coherence factors, see
Fig.~\ref{concurrence1}.

Bogoliubov showed \cite{Schrieff} that   natural excitations in the
superconducting state are a linear combination of particle and hole
excitations with the coherence factors $u_{\bk}$ and
  $v_{\bk}$. They  describe the unitary transformation from
  particle and hole operators to quasiparticles that are:
  \beqa
  \label{eq1}
  \gamma_{\bk, \uparrow} = u_{\bk}c_{\bk,
  \uparrow} + v_{\bk} c^\dag_{-\bk, \downarrow},
  \eeqa
  with the constraint   $ |u_{\bk}|^2+
  |v_{\bk}|^2
  = 1$ for any $\bk$ (normalization)

Constraint is
 satisfied with the choice
  \beqa
 |u_{k} |^2 = \sin^2
 \Theta_{k}, \nonumber \\
|v_{k}|^2   =   \cos^2 \Theta_{k}, \label{EQ:SCangle2}
 \eeqa
and  quantity
 \beqa
  \Theta_{k} =
  \arctan(\Huge(\frac{|u_k|^2}{|v_k|^2}\Huge)^{1/2})
  \label{EQ:SCangle}
  \eeqa
  Thus defined Bogoliubov angle $\Theta_{k}$
   is naturally related angle introduced by Anderson \cite{pwa58}.

To make a connection to ARPES we use the identity \cite{Schrieff}
\beqa A(k,\omega
>0)  = |u(\bf{k})|^2 \delta (\omega -
E_k),\nonumber\\
  A(k,\omega <0) = |v(\bf{k})|^2 \delta (\omega + E_k)
\label{EQ:SCangle4} \eeqa

Thus we have finally the main result to be used in ARPES \beqa
\Theta_{k} =
  \arctan ( \frac{A(\bk, \omega >0)}{A(\bk,\omega <0)})^{1/2}.
  \label{EQ:SCangle3}
  \eeqa
Experimentally $A(\bk, \omega >0),A(\bk,\omega <0)$ can be
determined using ARPES intensity at positive (above $E_F$) and
negative (below $E_F$) energy.

  \begin{figure}[htb]
\begin{center}
\includegraphics[width=2 in]{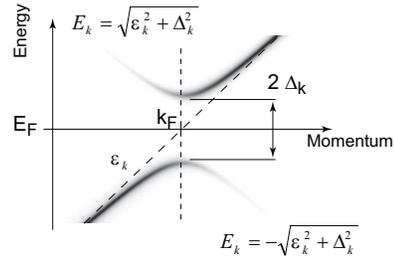}
\caption{\label{PH} Particle hole admixture in conventional BCS
superconductor as a function of energy is shown. Experimental
challenge is to find out BA as a function along the momentum on FS
in real high $T_c$ materials.}
\end{center}
\end{figure}

$\Theta_{k}$ is  the central quantity we are interested in and we
define it as
  a Bogoliubov angle.
 It represents a local mixture between particle and hole
 excitations for an eigenstate $n$ ( momentum  $k$ eigenstate in this context). For example, for $\Theta_{k} = 0$
 the Bogoliubov excitation will be a hole. In the opposite case of  $\Theta_{k} = \pi/2$ quasiparticle
 is essentially an electron. The angle that corresponds to the strongest admixture between particle and holes
 is  $\Theta_{k} = \pi/4 = 45^\circ$.



 ii) {\em ARPES and BA analysis}. We will now demonstrate how one can extract BA from ARPES.
 Although ARPES probes mostly the occupied portion of the single-particle spectral function \cite{Andrea03} (i.e. states
 below $E_F$, $\omega <0$) due to the Fermi-Dirac function $f(\omega, T$) cut-off near $E_F$, it is still possible to obtain some
 information about the states above $E_F$. Because of the high $T_c$ of some superconducting cuprates, ARPES measurements
 could be performed in the superconducting state at a relatively high temperature; such that the upper branch of the Bogoliubov
 quasiparticle dispersion near the nodal region, where the gap is smaller, could lay within the energy range where the value
 of $f(\omega, T)$ still appreciably differs from zero. In this situation, the Bogoliubov band dispersion above $E_F$ could be
 seen in the raw spectra allowing a further analysis of its property \cite{Matsui03_Bogoliubov}. In a recent study of a
 high-$T_c$ cuprate, Bi2212, the temperature dependence of the Bogoliubov dispersion were measured revealing a sudden onset of
 the superconducting gap at $T_c$ near the nodal region \cite{Lee07}. In this section, the Bogoliubov angle analysis
  was applied near the nodal region to demonstrate the concepts aforementioned. The experimental details of the data
  presented here can be found in Ref. \cite{Lee07}.

\begin{figure} [t]
\includegraphics [width=3.25 in]{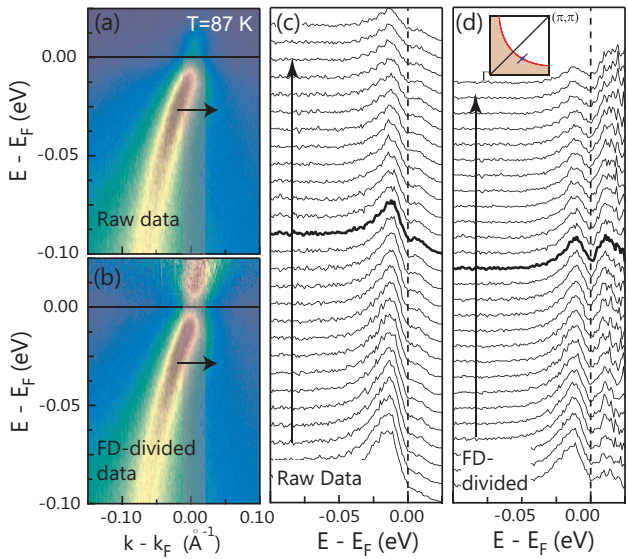}
\caption{(Color online)(a) The false-color plot of raw ARPES spectra
along the cut indicated in the inset of (d) in unit of the detector
angle. The data were taken on a slightly underdoped Bi2212 single
crystal with $T_C$ = 92 K at a temperature of 87 K using 7 eV
photons to excite the photoelectrons. The total energy resolution is
set to 3.2 meV. (b) Fermi-Dirac function divided spectrum of (a).
(c) EDCs stack plot of the raw spectra within the shaded area of
(a). (d) EDC stack plot of the FD-divided spectra within the same
shade area as also indicated in (b). The thick solid curve is the
EDC where $|u_k|^2 \sim |v_k|^2$ and could be defined as the Fermi
crossing point, $k_F$.} \label{Fig:imagesAndEDCs}
\end{figure}

Fig. \ref{Fig:imagesAndEDCs}(a) demonstrates a false color plot of
raw ARPES spectra along the cut position indicated in the inset of
(d) at 87 K. In addition to the high intensity region below the
Fermi energy (the occupied band dispersion), there is also a
less-bright region above $E_F$, which is the thermally populated
Bogoliubov band above $E_F$. The raw energy distribution curves
(EDCs)  in the region where the Bogoliubov dispersion is visible are
displayed in Fig. \ref{Fig:imagesAndEDCs} (c). A small peak above
$E_F$ representing the upper branch of Bogoliubov dispersion can be
clearly seen near the Fermi crossing momentum $k_F$, where the gap
is minimal. This small peak above $E_F$ becomes less pronounced when
moving away from $k_F$ because the peak position of the Bogoliubov
dispersion is moving away from $E_F$ (see also Fig. {\ref{PH}} ) ,
thus it is no longer able to be thermally populated at this
temperature.

\begin{figure} [t]
\includegraphics [width=3.25 in]{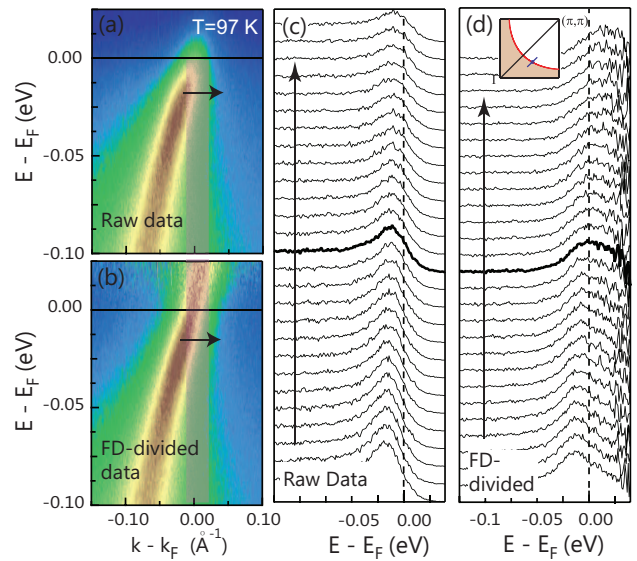}
\caption{ (Color online) Spectra taken at a temperature of T =97 K
(above $T_c$)
 at the same momentum position as that shown in Fig. \ref{Fig:imagesAndEDCs}. }
 \label{Fig:imagesAndEDCs_normal state}
\end{figure}

To illuminate these small features above $E_F$, the ARPES spectrum
is divided by an effective Fermi-Dirac function, which is generated
by convolving $f(\omega, T=87K$) with 3.2 meV instrument resolution
via a Gaussian convolution. The FD-divided ARPES spectrum image is
shown in Fig. \ref{Fig:imagesAndEDCs} (b). The intensity break near
the $E_F$ vividly demonstrates the existence of a gap and two
branches of dispersion centered at   $E_F$, as expected for the
Bogoliubov quasiparticle dispersion of a superconductor. The EDCs
within the shaded area are plotted in Fig.
\ref{Fig:imagesAndEDCs}(d). Two peaks in each EDC can now be clearly
seen exhibiting a evolution of the relative peak height at different
momentum positions due to the coherence factors. Before reaching
$k_F$ (EDCs below the thick solid curve), the peak below $E_F$ has a
higher intensity than that above $E_F$. After passing $k_F$ (EDCs
above the thick solid curves), the relation reverses; the peak above
$E_F$ now has  a higher intensity than that below $E_F$. This cross
over behavior near $k_F$ is also known to be a characteristic of the
Bogoliubov quasiparticle dispersion of a superconductor. Since the
peak intensity are related to the coherence factors $|u_k|^2$ and
$|v_k|^2$, they could be used for the Bogoliubov angle analysis
\cite{Matsui03_Bogoliubov}.

\begin{figure} [t]
\includegraphics [width=3.25 in]{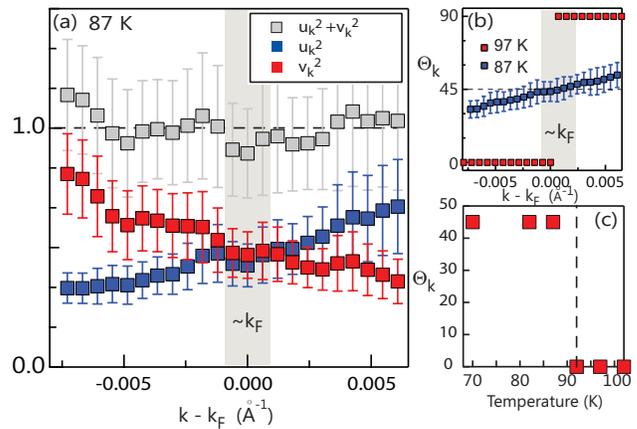}
\caption{ (Color online) (a) The extracted $|u_k|^2$ and $|v_k|^2$
at 87 K. (b) The Bogoliubov angle $\Theta_k$ near the Fermi crossing
point at 87 K. (c) The temperature dependence of $\Theta_k$ at
$k_F$. The error bars are estimated by the 99.7\% confidence
interval of the fitting parameters.} \label{Fig:BA}
\end{figure}

To extract the peak intensity, we simply fit the FD-divided EDCs
with two equal width Lorentizans in a narrow energy widow, ranging
from -20 meV to 20 meV, in which the signal has not yet completely
masked by the amplified noise due to the Fermi function division.
This noise amplification at higher energy above $E_F$ is also the
main reason why we use the peak intensity for the BA analysis,
instead of using peak area. We also note that this Lorentizan
fitting is primarily used for obtaining the peak height, not for
achieving a good fit to the spectrum line shape. Fig. \ref{Fig:BA}
(a) shows the extracted peak heights, which are normalized by the
average sum of the two peaks in each FD-divided EDC within this
momentum region. The normalized peak hight of the peak below $E_F$
is assigned a weight  $|v_k|^2$, while that of the peak above $E_F$
is assigned a weight $|u_k|^2$. Compared to Fig. 2(a) and (b), the
$|u_k|^2$ and $|v_k|^2$ extracted form the data are qualitatively
consistent with what is expected in the conventional BCS
superconductor. We also note that the sum of the extracted $|u_k|^2$
and $|v_k|^2$ is a constant within the experimental error bars. This
suggests that the normalization condition of $|u_k|^2 + |v_k|^2 = 1$
is satisfied in this momentum region, as a conventional BCS
superconductor does. We also note that this analysis is fully
consistent with an earlier work on a different cuprate
\cite{Matsui03_Bogoliubov}.

The Bogoliubov angle $\Theta_k$ is displayed in Fig. \ref{Fig:BA}
(b), which is calculated using Eq. 6. The $\Theta_k$ increase
monotonically across the Fermi crossing point $k_F$ suggesting a
continuously evolution of the particle and hole mixing within this
momentum window. Furthermore, $\Theta_k=\pi/4$ at $k_F$ within the
error bar of our experimental data. This confirms that the particle
and hole mix equally at $k_F$ as expected for a superconductor.

Situation is very different, however, at $T>T_c$.  No Bogoliubov
quasiparticle dispersion, nor a spectral gap at $E_F$ can be
resolved in this momentum region, as demonstrated in Fig.
\ref{Fig:imagesAndEDCs_normal state} (see also Ref. \cite{Lee07}).
There is only one peak can be identified in the Fermi function
divided EDCs, which disperses across $E_F$ (Fig.
\ref{Fig:imagesAndEDCs_normal state} (b) and (d)) suggesting there
is no gap in the spectral. To generalize Bogoliubov angle to this
situation, we define ($u_k, v_k$) = (1,0) when the quasiparticle
peak in the FD-divided EDC is below or at $E_F$, and ($u_k, v_k$) =
(0,1) when the peak of the FD-divided EDC is above $E_F$. This will
induce an angle jump from 0 to 90 deg near the $k_F$ as shown in
Fig. \ref{Fig:BA} (b). Thus, $\Theta_k$ equals to zero at $k_F$ and
a sudden jump across the $K_F$ suggests an absence of the particle
and hole mixing above $T_c$ at this momentum position. The
$\Theta_k$ at $k_F$ at several different temperatures is shown in
Fig. \ref{Fig:BA}(c). Here we used notations that the Bogoliubov
angle is zero by taking its value for filled states.  The behavior
of the electron at this momentum position (near nodal region)
appears to be very conventional, despite the existence of a
pseudogap near the antinodal region, see also  \cite{Lee07} and the
reference therein.

\begin{figure} [t]
\includegraphics [width=3.0 in]{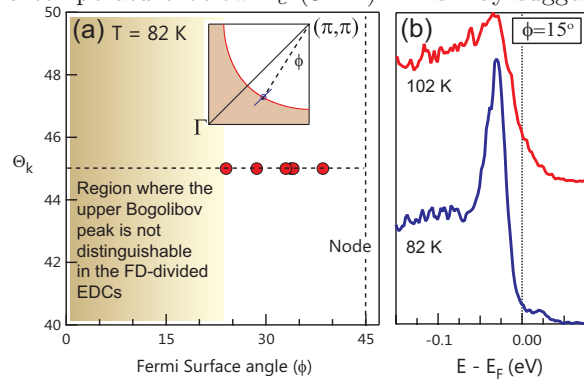}
\caption{(Color online) (a) The Bogoliubov angle at a temperature 82
K along the Fermi surface where the Bogoliubov peak can be
unambiguously distinguished from the noise level in the FD-divided
spectrum. (b) Raw  EDC at $\phi = 15^\circ$ at temperatures above
(102 K, red curve) and below $T_c$ (82K, blue curve). At this
momentum position, a gap (pseudogap) still exists  in the spectrum
of 102 K.} \label{Fig:BA_K_DEP}
\end{figure}

The BA at several different positions near the nodal region on the
Fermi surface at a temperature of 82 K is shown in Fig.
\ref{Fig:BA_K_DEP}(a). The BA is found to be $\pi/4$ suggesting
again that the particle and hole mix equally on the Fermi surface at
least near the nodal region. For the region in the shaded area
indicated in Fig. \ref{Fig:BA_K_DEP}(a), the upper branches of the
Bogoliubov band disperses too far way from the $E_F$ and become less
accessible by the thermal energy at this temperature. We note that
although the feature of Bogoliubov peak above $E_F$ can still be
identified in the raw EDCs near $k_F$ up to $\phi=15^\circ$, the
Bogoliubov peak above $E_F$ in the FD-divided spectrum is too noisy
to be useful for the BA analysis. Therefore, we can't obtain any
conclusive information about the BA for the momentum positions
within this shaded area of Fig. \ref{Fig:BA_K_DEP}. We also remark
that at a temperature above $T_c$, Fig. \ref{Fig:BA_K_DEP} (b), the
Bogoliubov quasiparticle  peak above $E_F$ has not yet been resolved
in the pseudogap state  at this momentum region, even through it is
well defined  at a temperature below $T_c$ (82 K). This may suggest
a qualitative difference of the particle-hole mixing in the
pseudogap state from that of the superconducting state. However, we
could not rule out the possibility that the absence of a Bogoliubov
quasiparticle dispersion feature above $T_C$ at this momentum region
region is due to the significant broadening of the peak in the
spectrum. Further study is needed to clarify this issue.

In conclusion, we have introduced a new spectroscopic measure,
Bogoliubov angle $\Theta(\bk)$. BA can be extracted from existing
ARPES data. This measure allows one to image  particle-hole
admixture in the superconducting state.

The ideas presented here are quite general
 and are applicable to a variety of superconductors, including conventional
 superconductors. One can investigate Bogoliubov angle in a variety of states,
 including vortex state and normal state  with superconducting correlations,
  e.g. so called pseudogap (PG) state.   As a future application of these ideas we suggest
  using Bogoliubov angle to
identify how robust the particle-hole mixture is in the normal state
of cuprates. At present stage we do not have enough resolution to
perform this analysis. Another interesting question is the  BA
behavior with
 temperature. Answers to these questions will shed light on the nature of PG state and
  would allow us to differentiate between
different scenarios of PG state, e.g. flux phases \cite{Wen} and D
wave density wave (DDW) \cite{Chakravarty2000}.

We are grateful to I. Grigorenko, J.C. Davis, A. Yazdani, N.
Nagaosa, M. Randeria and O. Fischer for useful discussions. This
work has been supported by US DOE, OBES, LDRD.

\setcounter{equation}{0}
\renewcommand{\theequation}{A-\arabic{equation}}
{99}


\end{document}